\documentclass[pra,amssymb,amsfonts,amsmath,superscriptaddress,showpacs]{revtex4}

\newcommand{\tr}{\mbox{tr}}
\newcommand{\bra}[1]{\ensuremath{\langle #1 |}}
\newcommand{\ket}[1]{\ensuremath{| #1 \rangle}}
\newcommand{\bk}[2]{\ensuremath{\langle #1 | #2 \rangle}}
\newcommand{\kb}[2]{\ensuremath{| #1 \rangle\!\langle #2 |}}

\begin{document}

\title{Universal nonlinear entanglement witnesses}
\author{Marcin Kotowski}
\affiliation{\it College of Inter-Faculty Individual Studies in Mathematics and Natural Sciences, Warsaw University, Warszawa, Poland.}
\affiliation{Center for Theoretical Physics, Polish Academy of Sciences,
Aleja Lotnik{\'o}w 32/44, 02-668 Warszawa, Poland}
\author{Micha{\l} Kotowski}
\affiliation{\it College of Inter-Faculty Individual Studies in Mathematics and Natural Sciences, Warsaw University, Warszawa, Poland.}
\affiliation{Center for Theoretical Physics, Polish Academy of Sciences,
Aleja Lotnik{\'o}w 32/44, 02-668 Warszawa, Poland}
\author{Marek Ku\'s}
\affiliation{Center for Theoretical Physics, Polish Academy of Sciences,
Aleja Lotnik{\'o}w 32/44, 02-668 Warszawa, Poland}

\begin{abstract}
We give a universal recipe for constructing nonlinear entanglement witnesses
able to detect non-classical correlations in arbitrary systems of
distinguishable and/or identical particles for an arbitrary number of
constituents. The constructed witnesses are expressed in terms of expectation
values of observables. As such they are, at least in principle, measurable in
experiments.
\end{abstract}

\date{\today}
\maketitle

\section{Introduction}
Nonclassical correlations among subsystems of a composite quantum system,
known as entanglement, can be easily characterized mathematically but, at
least in the case of mixed states, in a rather ineffective way. In general,
given a mixed state of a composite system it is hard to decide whether it is
separable (nonentangled) or not with respect to a given partition of the
whole system into subsystems. In the case of many (more than two) subsystems
even determination of separability of a pure state might pose a computational
problem.

It is thus desirable to construct a `measure of entanglement', i.e.\ a
function from the set of quantum states into real numbers which vanishes on
separable states and takes nonzero, say positive, values for the nonseparable
ones. If we impose some further, natural and reasonable conditions to be
fulfilled by such a function, we may obtain a useful, quantitative
characterization of `the amount of entanglement' in a given state. Evidently
such a quantitative measure is not unique \cite{bengtsson06} and the choice
of a particular measure is dictated by a particular case we want to analyze.
In the following we will mostly focus our attention on discriminating between
separable and nonseparable states. In terms of an entanglement measure it
means that we are interested only if it takes zero or non-zero value on a
state under investigation.

To be of practical use in experiments, an entanglement measure should be
measurable, i.e.\ it should be possible to design an experiment in which a
value of the measure for a particular state can be established. Since in
quantum mechanics we can measure only observables, a measurable measure
should be given as an expectation value of a Hermitian operator calculated in
the investigated state. It is, however, straightforward to check that there
is no observable such that its expectation values vanish exactly on pure
separable states \cite{badziag02}. Such known measures of entanglement like,
e.g., Wootter's concurrence can be expressed in terms of expectation values
of \emph{antilinear} operators \cite{wootters98,uhlmann00}, and as such are
not directly measurable. It is however possible to find \emph{bilinear}
(i.e.\ acting on two copies of a state) Hermitian operators for which the
condition of vanishing expectation values are fulfilled only by pure
separable states
\cite{mkb04,mkb05,mckb05,mintert07,mintert07a,mintert07b,aolita06,%
aolita08,zhang08a}.

In the following we show how to construct such measures for pure states in
the two-partite as well as many-partite systems in an `algorithmic' way.
Moreover the constructed measures allow to estimate from below the amount of
entanglement for mixed states and thus provide some effective (although not
always decisive) criteria of entanglement. One of our main points is to
stress the unifying character of the presented approach, allowing for a
uniform treatment of arbitrary number of distinguishable and identical
particles (bosons and fermions).

\section{Pure separable states}
As it is customary in quantum information theory, for which entanglement is
one of the most important resources, we will investigate quantum systems in
finite-dimensional Hilbert spaces (one should think about various spin or
spin-like system, multilevel atoms etc.). Thus with a quantum system we
associate a Hilbert space $\mathcal{H}$ isomorphic to $\mathbb{C}^N$.
Customarily, vectors from $\mathcal{H}$ are called (pure) states of the
system. One should, however, keep in mind that in order to give a
probabilistic interpretation to amplitudes we normalize states to unit norm
(length) and moreover disregard an irrelevant phase of the vector. It is thus
more appropriate to think of states as points in the projective space
$\mathbb{P}(\mathcal{H})$. Alternatively and equivalently, it is convenient
to treat a pure state as a one-dimensional projection operator
$P_\psi:=\kb{\psi}{\psi}/\bk{\psi}{\psi}$, freeing ourselves from the
normalization and phase problems and unifying the treatment of pure and mixed
states by identifying both with positive-definite operators on $\mathcal{H}$
with unit trace, where pure states are distinguished by $\rho^2=\rho$
exhibiting their projective character.

For composed systems the Hilbert space $\mathcal{H}_{comp}$ is a tensor
product of the Hilbert spaces of subsystems,
$\mathcal{H}_{comp}=\mathcal{H}_1\otimes
\mathcal{H}_2\otimes\cdots\otimes\mathcal{H}_n$. The states represented by
simple tensors, i.e.\ $\mathcal{H}_{comp}\ni\ket{\Phi}=\ket{\phi_1}\otimes
\ket{\phi_2}\otimes\cdots\ket{\phi_n}$ are called pure separable states.
Identifying a pure state with a one-dimensional projection operator we call
such a projection separable if it projects on the direction of a separable
vector. We then define separable mixed states as the convex hull of the
separable projections. It is easily seen that this definition of separable
states coincides with the commonly adopted one - a state is separable if it
is a combination of product states with positive coefficients,
$\rho=\sum_ip_i\rho^{1}_i\otimes\rho^{2}_i\otimes \cdots\otimes\rho^{n}_i$,
$\rho^{k}_i$ - a state on $\mathcal{H}_i$, $p_i>0$. The states that are not
separable are called entangled. From the point of quantum information
technology the really interesting states, useful, e.g., in secure and more
effective transmission of information, are entangled states, and that is why
effective discrimination between them and separable ones is of importance
(see, e.g., \cite{horodecki09}).

In the case of bipartite systems an arbitrary pure state can be represented
by a vector
\begin{equation}\label{s1}
\ket{\psi}=\sum_{i,j}c_{ij}\ket{e_i}\otimes\ket{f_j},
\end{equation}
where $\{\ket{e_i}\}_{i=1}^N$ and $\{\ket{f_j}\}_{j=1}^M$ are some
orthonormal bases in $\mathcal{H}_1$ (of dimension $N$) and $\mathcal{H}_2$
(of dimension $M$), and $c=(c_{kl})$ is a complex matrix. By unitary changes
of the bases $\ket{e_i}=\sum_kU_{ki}\ket{e_k^\prime}$,
$\ket{f_j}=\sum_lV_{lj}\ket{f_l^\prime}$, which amounts to transformation
$c\mapsto UcV^T:=c^\prime$, with $^T$ standing for transposition, one can
bring (\ref{s1}) to its Schmidt form,
\begin{equation}\label{s2}
\ket{\psi}=\sum_k\lambda_k\ket{e_k^\prime}\otimes\ket{f_k^\prime}, \quad
\lambda_k>0.
\end{equation}
A pure state is nonentangled if and only if only one of its Schmidt
coefficients $\lambda_k$ does not vanish. The coefficients $\lambda_k$ are
positive real numbers squares of which are non-zero eigenvalues of $c^\dagger
c$ (or, equivalently $cc^\dagger$), hence are easily calculable for a given
$\ket{\psi}$,

The definition of separability in terms of simple tensors lacks sense for
systems of identical particles when states must be symmetric or antisymmetric
with respect to relabeling of subsystems (particles). Indeed due to the
(anti)symmetrization nearly all state vectors do not have the product form;
the only exception is the state of bosons each occupying the same single
particle state). For example, the simplest state vector of two identical
fermions reads as
$\ket{\Phi}=\ket{\phi_1}\otimes\ket{\phi_2}-\ket{\phi_2}\otimes\ket{\phi_1}
=:\ket{\phi_1}\wedge\ket{\phi_2}$ which according to the usual definition is
entangled \footnote{Here and in the following we will not pay attention to
the normalization of vectors since it does not play any crucial role,
moreover is taken into account if we pass to the projective space.}.

In a way, states of identical particles exhibit some \textit{a priori}
correlations and only additional amount of correlation should be classified
as entanglement. Several, not necessarily equivalent ways were proposed to
identify and quantify this phenomenon. In \cite{schliemann01,sckll01} a
correlation measure for states of two undistinguishable fermions was
proposed. It is constructed in analogy with the distinguishable particles
case by employing algebraic properties of the coefficient matrix in the
expansion of a state in terms of basis states. To this end one observes that
a pure states of two undistinguishable fermions in an $n$-dimensional single
particle space $\mathcal{H}$ can be written as
\begin{equation}\label{purefermions}
\ket{w}=\sum_{i,j=1}^nw_{ij}f_i^\dagger f_j^\dagger\,\ket{0},
\end{equation}
where $w=(w_{ij})$ is a complex antisymmetric matrix fulfilling the
normalization condition $\tr(w^\dagger w)=\frac{1}{2}$. Here $f_i^\dagger$
are fermionic creation operators and $\ket{0}$ is the vacuum state. A unitary
transformation $U$ in the single particle space changes $w$ to
$w^\prime=UwU^T$. By choosing an appropriate unitary $U$ we can transform $w$
to its canonical, block-diagonal form:
\begin{equation}\label{purefermions1}
w=\mathrm{diag}[Z_1,\ldots,Z_r,Z_0], \quad
Z_i=\left[
\begin{array}{cc}
0 & z_i \\
-z_i & 0
\end{array}
\right], \quad z_i>0,
\end{equation}
where $2r$ is the rank of $w$ and $Z_0$ is the null matrix of dimension
$(n-2r)\times (n-2r)$ \cite{horn85}. The squares of the coefficients $z_i$
are eigenvalues of $ww^\dagger$, hence again are easily calculable for a
given $w$. A state is, by definition, nonentangled if only one of $z_i$ does
not vanish, i.e. only a single, elementary $2\times 2$ Slater determinant
appears in the canonical decomposition (\ref{purefermions1}). In other words
a state is nonentangled if it can be written as
$\ket{\psi}=f^{\prime\dagger}_1f^{\prime\dagger}_2\ket{0}$ where
$f^{\prime\dagger}_i\ket{0}$, $i=1,\ldots,n$ form an orthonormal basis in the
single particle Hilbert space, i.e.\ $\ket{\psi}$ is the antisymmetrization
of a product state. Measuring quantum correlations by the number of
nonvanishing terms when expanding a wave function in terms of elementary
Slater determinants was proposed earlier \cite{herbut87,grobe94}, employing
various quantitative characterizations of the number of non-zero terms.

The above idea was thoroughly investigated in
\cite{schliemann01,sckll01,eckert02}. Generalizing the two-particle case, the
relevant definition of pure nonentangled state of $n$ indistinguishable
fermions (we will use the notion of `nonentanglement' rather than
`separability' which in this context lacks its semantic sense), can be
shortly summarized as follows. A fermionic state is nonentangled if and only
if it can be written as $f_1^\dagger f_2^\dagger\cdots f_n^\dagger\ket{0}$,
i.e.\ it is the complete antisymmetrization of a product state. This
coincides with definitions proposed in \cite{li01,ghirardi02}, where
conclusions were reached by departing from slightly different starting
points.

Similar ideas can be applied to bosons \cite{paskauskas01,li01,eckert02}. A
general, two-particle state in an $n$-dimensional single particle space,
\begin{equation}\label{purebosons}
\ket{v}=\sum_{i,j=1}^nv_{ij}b_i^\dagger b_j^\dagger\,\ket{0},
\end{equation}
where $b_i^\dagger$ are boson creation operators and $v=(v_{ij})$ is a
complex symmetric matrix, can be transformed to a canonical form with
$v_{ij}=z_i\delta_{ij}$ by a unitary transformation in the single-particle
space, amounting to $v\mapsto UvU^T$ on the level of the coefficient matrix
\footnote{This algebraic fact is known as the Takagi factorization theorem,
see \cite{horn85}.}. Consequently a pure $n$-boson state is nonentangled if
it can be written as $\ket{v}=b_1^\dagger\cdots b_1^\dagger\ket{0}$. It
should be pointed that in \cite{li01} and \cite{ghirardi04} (see
also\cite{herbut01}), in contrast to \cite{paskauskas01,eckert02}, a slightly
broader definition of pure nonentangled boson states was proposed. In
addition to the identified above, as nonentangled are also treated states
which in some basis can be written as $b_1^\dagger\cdots b_N^\dagger\ket{0}$,
where all states $b_k\ket{0}$ are orthogonal. We would like to make two
comments concerning this point. Firstly, the extended class of states can be
also easily described using methods proposed in this paper (as orbits of
unitary groups - see below), although admittedly, characterization in terms
of vanishing expectation value of some bilinear operator needs more efforts
which we postpone to other occasion. Secondly, the real meaning of
entanglement becomes important in particular experiments. Whether it can be
exhibited (especially for indistinguishable particles) depends strongly of
what and how we measure states in an experiment \cite{tmkb-preprint}, thus,
e.g., the authors of \cite{paskauskas01} mention that states identified by Li
et al.\ \cite{li01} as nonentangled can be used for quantum teleportation,
and as such should be treated as quatally correlated (entangled). In the
present paper we would like to retain the more restrictive definition of
bosonic nonentanglement, leaving, as already stated, the more relaxed one for
the topic of future investigations.

We would like to conclude this section by stressing that there are no analogs
of the Schmidt and Takagi decompositions in multipartite cases, therefore
establishing nonentanglement demands considerably more elaborate methods
\cite{sckll01,eckert02}.

\section{Actions of unitary groups in composite systems spaces}

In order to achieve the goals of characterizing separability \textit{via}
observables let us look at the problem from a slightly different point of
view. For the moment we restrict the attention to bipartite systems. It is
obvious that separability of a pure state of distinguishable particles does
not change if we individually evolve each subsystem in a quantum-mechanically
allowed way i.e.\ \textit{via} a unitary transformation,
$\ket{\phi_1}\otimes\ket{\phi_2}\mapsto U_1\ket{\phi_1}\otimes
U_2\ket{\phi_2}$. In the case of indistinguishable particles the same is true
if we perform the same unitary transformation $U$ in each one-particle space,
$\ket{\phi_1}\wedge\ket{\phi_2}\mapsto U\ket{\phi_1}\wedge U\ket{\phi_2}$ or
$\ket{\phi_1}\vee\ket{\phi_2}\mapsto U\ket{\phi_1}\vee U\ket{\phi_2}$ (where
$\vee$ denotes the symmetrized tensor product). In fact we used this
invariance to transform pure states to their canonical Schmidt or Takagi
forms in the preceding section. We may assume that the matrices have
determinant one, so they are elements of special unitary groups, since the
phase of the state does not play any role. The action of the special unitary
group on vectors from a Hilbert space $\mathcal{H}$ translates in a natural
way to an action on the projective space $\mathbb{P}(\mathcal{H})$. If we
denote by $[v]$ the point in $\mathbb{P}(\mathcal{H})$ (the direction of
$\ket v$, or in other words the complex line through $\ket v$) we have, by
definition, $U[v]=[U\ket v]$. This action of a unitary group on
$\mathbb{P}(\mathcal{H})$ is transitive, i.e.\ any two points in
$\mathbb{P}(\mathcal{H})$ can be connected by some unitary transformation. A
straightforward conclusion is that any two nonentangled states in spaces of
states of bipartite systems are connected by the above described unitary
action of the direct product of two unitary groups in the case of
distinguishable particles and a single unitary group for fermions and bosons.
In more technical terms nonentangled state in all cases form a single orbit
of an action of the group $SU(N)\times SU(M)$ in
$\mathbb{P}(\mathcal{H}_1\otimes\mathcal{H}_2)$ for distinguishable particles
or $SU(N)$ in the case of bosons or fermions in
$\mathbb{P}(\mathcal{H}\vee\mathcal{H})$ or
$\mathbb{P}(\mathcal{H}\wedge\mathcal{H})$, respectively.

In order to identify uniquely the orbit in question we have to make a short
excursion in the theory of representation of semisimple Lie groups (the
special unitary group $SU(N)$ is simple and the direct product of simple
groups, such as $SU(N_1)\times SU(N_2)\times\cdots\times SU(N_k)$ is
semisimple \cite{hall03}). Let $K$ be a semisimple real group and $G$ its
complexification, and denote by $\mathfrak{g}$ the Lie algebra of $G$. For
$K=SU(N)$ we have $G=SL(N,\mathbb{C})$ (special complex linear group in $N$
dimensions) and $\mathfrak{g}=\mathfrak{sl}_N(\mathbb{C})$ (the algebra of
complex $N\times N$ matrices with vanishing trace). As a basis in
$\mathfrak{sl}_N$ we can choose $N-1$ independent traceless diagonal matrices
$H_k$ which span its maximal commutative subalgebra and the matrices $X_{ij}$
having a single nonvanishing element on in the $(i,j)$ position. The
commutation relations among $H_k$ and $X_{ij}$ read as:
$[H_k,X_{ij}]=\alpha_{ij}(H_k)X_{ij}$ where $\alpha_{ij}$ is some linear
function on the set of diagonal matrices. An analogous construction exists
for an arbitrary semisimple complex Lie algebra. We can distinguish in it a
maximal commutative subalgebra $\mathfrak{t}$ of dimension $r$ (called the
rank of the algebra) and one-dimensional subspaces (root spaces)
$\mathfrak{g}_\alpha$ (spanned by $X_\alpha$) such that
\begin{equation}\label{roots}
[H,X]=\alpha(H)X, \quad H\in\mathfrak{t},\quad X\in\mathfrak{g}_\alpha.
\end{equation}
If we choose a basis $\{H_k\}_{k=1}^r$ in $\mathfrak{t}$, the algebra is
uniquely determined by the set of vectors
$\boldsymbol\alpha=(\alpha(H_1),\ldots,\alpha(H_r))$ where $\alpha$ runs over
all different root spaces. The linear form $\alpha$ is called a root of
$\mathfrak{g}$, and the element $X_\alpha$ corresponding to the root $\alpha$
is a root vector. There always exists a natural symmetry: to each $\alpha$
corresponds $-\alpha$, as in the described above case of
$\mathfrak{sl}_N(\mathbb{C})$ treated as the algebra of $N\times N$ traceless
matrices, to each $X_{ij}$ with, say, $i<j$, i.e.\ with a single nonvanishing
element in the upper-right triangle, there corresponds $X_{ji}$ living in the
lower-left triangle of the matrix for which the commutators with $H_k$ have
the opposite sign to those of the commutators of $X_{ij}$. It means that to
characterize an algebra we need only half of the root vectors - the
`positive' ones corresponding e.g., to upper triangular matrices which we
will call positive root vectors and the corresponding roots $\alpha$ - the
positive roots ($\alpha>0$).

The groups $K$ and $G$, as well as the Lie algebra $\mathfrak{g}$ can be
represented irreducibly in spaces of different dimensions, i.e.\ to each
element of the group or algebra there corresponds a linear operator acting in
some Hilbert space, say, $\mathbb{C}^M$, such that the group multiplication
and the Lie bracket (commutator) in the algebra are preserved. Hence,
denoting by $\pi(X)$ the representative of the Lie algebra element $X$ we
have from (\ref{roots})
\begin{equation}\label{roots1}
[\pi(H),\pi(X)]=\alpha(H)\pi(X), \quad H\in\mathfrak{t},
\quad X\in\mathfrak{g}_\alpha.
\end{equation}
Since the operators $\{H_k\}_{k=1}^r$ commute, the same is true for their
representatives $\{\pi(H_k)\}_{k=1}^r$. It follows that $\pi(H_k)$ have
common eigenvectors. For each irreducible representation there exists a
unique (up to a multiplicative constant) eigenvector $\ket{v_{max}}$ of all
$\pi(H_k)$ which is annihilated by all the representatives of the positive
roots
\begin{equation}\label{maxweight}
\pi(H_k)\ket{v_{max}}=\lambda_k\ket{v_{max}},\quad k=1,\ldots,r,
\quad \pi(X_\alpha)\ket{v_{max}},\quad \alpha>0.
\end{equation}
The vector $\ket{v_{max}}$ is called the maximal weight-vector. An
irreducible representation of $\mathfrak{g}$ (and, in consequence, of $G$ and
$K$) is uniquely determined by the eigenvalues $\lambda_k$ which we cast in a
$r$-component vector $\boldsymbol{\lambda}=(\lambda_1,\ldots,\lambda_r)$.

It is now easy to identify the orbit of $SU(N)\times SU(M)$ in
$\mathbb{P}(\mathcal{H}_1\otimes\mathcal{H}_2)$ of the nonentangled states as
the orbit through the maximal weight vector for the representation of
$SU(N)\times SU(M)$ in
$\mathcal{H}_1\otimes\mathcal{H}_2\simeq\mathbb{C}^N\otimes\mathbb{C}^M$.
Indeed, since the action of $SU(N)$ on $\mathbb{P}(\mathbb{C}^M)$ is
transitive, each point $[v]$ is on the orbit $\{[U\ket{v_{max}}]: U\in
SU(N)\}$. Thus each nonentangled state is on the orbit of $SU(N)\times SU(M)$
of the state $[{w_{max}}]:=[{v_{max}}\otimes{u_{max}}]$ where $\ket{v_{max}}$
and $\ket{u_{max}}$ are the highest-weight vectors of the representations of
$SU(N)$ in $\mathbb{C}^N$ and $SU(M)$ in $\mathbb{C}^M$. On the other hand,
it is easy tho see that $\ket{w_{max}}$ is exactly the highest-weight vector
of the irreducible representation $SU(N)\times SU(M)$ in
$\mathcal{H}_1\otimes\mathcal{H}_2\simeq\mathbb{C}^N\otimes\mathbb{C}^M$.
Similar considerations show that analogous statements are true in the cases
of many particles with and/or without assumptions about their
distinguishability.

Summarizing, the nonentangled states form the orbit in the projective space
$\mathbb{P}(\mathcal{H}_{comp})$ through the highest-weight vector of an
irreducible representation of a unitary group $K$ in $\mathcal{H}_{comp}$.
For $n$-distinguishable particles $K=SU(N_1)\times\cdots\times SU(N_n)$ and
$\mathcal{H}_{comp}=\mathcal{H}_1\otimes\cdots\otimes\mathcal{H}_n$,
$\dim\mathcal{H}_k=N_k$, whereas for indistinguishable particles $K=SU(N)$
and $\mathcal{H}_{comp}=\mathcal{H}\vee\cdots\vee\mathcal{H}$ (for bosons) or
$\mathcal{H}_{comp}=\mathcal{H}\wedge\cdots\wedge\mathcal{H}$ (for fermions),
$\dim\mathcal{H}=N$.

There exists a nice and simple method of characterizing the orbit through
highest weight vector of an irreducible representation of a semisimple group
\cite{lichtenstein82}. First we define a second-order operator
\begin{equation}\label{casimir}
C_2:=\sum_{\alpha>0}\Big(\pi(X_\alpha) \pi(X_{-\alpha})+\pi(X_{-\alpha})
\pi(X_{\alpha})\Big)
+\sum_{i=1}^r\pi(H_i)\pi(H_i).
\end{equation}
It is called the (second-order) Casimir operator; using (\ref{roots1}) one
shows that it commutes with all operators of the representation of the
algebra, and thus (for an irreducible representation) is proportional to the
identity operator $I$ \footnote{We use the same symbol $I$ to denote the
identity operator in an arbitrary space. From the context it is usually
obvious in which space $I$ actually acts.}. In fact one can prove that
$C_2=\langle\boldsymbol\lambda,\boldsymbol\lambda +2\boldsymbol\delta\rangle
I$, where $\boldsymbol\delta=\frac{1}{2}\sum_{\alpha>0}\boldsymbol{\alpha}$
is the half-sum of the positive roots and $\langle\, ,\rangle$ is the
Euclidean scalar product in the space of $r$-dimensional vectors to which
$\boldsymbol\lambda,\boldsymbol\alpha$, and $\boldsymbol\delta$ belong
\cite{barut80}.

A vector $\ket{\psi}$ belongs to the highest-weight orbit of the irreducible
representation $\pi$ with the highest weight $\boldsymbol\lambda$ of a
semisimple group $G$ if and only if \cite{lichtenstein82}
\begin{equation}\label{orbit}
L\ket\psi\otimes\ket\psi=
\langle 2\boldsymbol\lambda+2\boldsymbol\delta,2\boldsymbol\lambda\rangle
\ket\psi\otimes\ket\psi,
\end{equation}
where
\begin{equation}\label{L}
L=C_2\otimes I+I\otimes C_2
+2\sum_{\alpha>0}\Big(\pi(X_\alpha)\otimes\pi(X_{-\alpha})
+\pi(X_{-\alpha})\otimes\pi(X_{\alpha})\Big)+
2\sum_{i=1}^r \pi(H_i)\otimes\pi(H_i).
\end{equation}

Relevant properties of the operator $L$ for the cases of two distinguishable
particles, fermions, and bosons are calculated in the Appendix. For general
cases of $n$ particles, distinguishable or not, similar calculations can be
also performed (with considerably more effort).

\section{Measurable entanglement measures}
\label{sec:measures}

The operator $L$ is Hermitian and positive semidefinite and its largest
eigenvalue equals $l_{max}=\langle
2\boldsymbol\lambda+2\boldsymbol\delta,2\boldsymbol\lambda\rangle$. Hence
$A:=l_{max}I-L$ is positive semidefinite and its expectation value
$\bra{\psi}\otimes\bra{\psi}A\ket{\psi}\otimes\ket{\psi}$ vanishes exactly
for a nonentangled $\ket\psi$. This is the desired characterization of the
nonentangled states in terms of an expectation value of an observable, i.e.
$A$ can be regarded as a `nonlinear' entanglement witness.

The operator $A$ acts in the tensor product
$\mathcal{H}_{comp}\otimes\mathcal{H}_{comp}$. There is a natural isomorphism
(the Jamio{\l}kowski isomorphism \cite{jamiolkowski72}) between the set of
operators on $\mathcal{H}_{comp}\otimes\mathcal{H}_{comp}$ (such as the
operator $A$) and the set of operators acting on density matrices (or
generally linear operators) on $\mathcal{H}_{comp}$ given by
\begin{equation}\label{jam}
\Lambda(\rho)=\tr_1\Big((\rho^T\otimes I) A\Big),
\end{equation}
where $^T$ denotes the transposition of a matrix and $\tr_1$ is the trace
over the first factor of the tensor product
$\mathcal{H}_{comp}\otimes\mathcal{H}_{comp}$. The most important feature of
(\ref{jam}) is, for our purposes, that for positively semidefinite $A$ the
operator $\Lambda$ is completely positive. The actual definition of complete
positiveness is not important here - what is crucial is that it is equivalent
to the fact that the action of $\Lambda$ can be expressed in the so-called
Kraus form \cite{bengtsson06}
\begin{equation}\label{krauss}
\Lambda(\rho)=\sum_{\mu=1}^sT_\mu\rho T_\mu^\dagger.
\end{equation}
The operators $T_\mu$ can be expressed in terms of eigenvectors of $A$. Let
\begin{equation}\label{spA}
A=\sum_{\mu=1}^s\nu_\mu\kb{v_\mu}{v_\mu}
\end{equation}
be the spectral decomposition of $A$ and $s$ the rank of $A$. Since $A$ is
positive semidefinite we have $\nu_\mu>0$ and defining
$\ket{w_\mu}=\sqrt{\nu_\mu}\ket{v_\mu}$ we obtain
\begin{equation}\label{spA1}
A=\sum_{\mu=1}^s\kb{w_\mu}{w_\mu}.
\end{equation}
Let us now choose in $\mathcal{H}_{comp}$ an orthonormal basis $\ket{e_i}$,
$i=1,\ldots,N=\dim\mathcal{H}_{comp}$, and expand the unnormalized
eigenvectors $\ket{w_\mu}$ of $A$ in this basis,
\begin{equation}\label{wmu}
\ket{w_\mu}=\sum_{i,j=1}^Nw^{(\mu)}_{ij}\ket{e_i}\otimes\ket{e_j}
\end{equation}
It is now easy to find that
\begin{equation}\label{T}
T_\mu=\Big(\sum_{k=1}^N\bra{e_k}\otimes\bra{e_k}\otimes I\Big)
\Big(I\otimes\big(\sum_{i,j=1}^Nw^{(\mu)}_{ij}\ket{e_i}
\otimes\ket{e_j}\big)\Big)=\sum_{i,j=1}^Nw^{(\mu)}_{ij}\kb{e_i}{e_j}.
\end{equation}
Another straightforward calculation reveals the matrix elements of $A$ in
terms of $T_\mu$ \cite{kb09}
\begin{equation}\label{kb}
 \bra{\psi_2\otimes\psi_4}A\ket{\psi_1\otimes\psi_3}=\sum_{\mu=1}^s
\bra{\psi_2}T_\mu\ket{\psi_4^\ast}\bra{\psi_1^\ast}
T_\mu^\dagger\ket{\psi_3}.
\end{equation}
In particular, with the help of the expectation value of $A$ we can construct
a pure-state entanglement measure (generalized concurrence)
\begin{equation}\label{psiApsi}
c_A(\psi):=\bra{\psi\otimes\psi}A\ket{\psi\otimes\psi}^{1/2}=
\Big(\sum_{\mu=1}^s|\bra{\psi}T_\mu\ket{\psi^\ast}|^2\Big)^{1/2},
\end{equation}
with the property $c_A(\alpha\psi)=|\alpha|^2c_A(\psi)$ \cite{mkb04,mckb05}.

Since the Kraus operators $T_\mu$ are calculable from the eigenvectors of $A$
and the latter are the same for $L$ which differs from $A$ by a multiple of
the identity operator, we can find them by decomposing the image of $L$ under
the Jamio{\l}kowki isomorphism into the Kraus form. When calculating $T_\mu$
from eigenvectors of $L$ we should disregard those which correspond to the
maximal eigenvalue $l_{max}$ of $L$ since they belong to the zero eigenvalue
of $A$ and as such do not contribute to the Kraus decomposition of its image
under the Jamio{\l}kowski isomorphism [cf.\ (\ref{spA},\ref{spA1})]. In fact,
we may disregard even more of the Kraus operators $T_\mu$ calculated from
$L$. Indeed due to the appearance of the complex conjugate vector
$\ket{\psi^*}$ in (\ref{psiApsi}), the terms $\bra{\psi}T_\mu\ket{\psi^\ast}$
vanish for antisymmetric $T_\mu$, hence for our purposes only symmetric
$T_\mu$ are of interest when determining $c_A$. For example, as shown in the
Appendix, only the subspace $\mathcal{H}_S^1$ of
$\mathcal{H}_{comp}\otimes\mathcal{H}_{comp}=
\mathcal{H}_1\otimes\mathcal{H}_2\otimes\mathcal{H}_1\otimes\mathcal{H}_2$
obtained by antisymmetrizing separately the $\mathcal{H}_1$ factors and the
$\mathcal{H}_2$ ones produces relevant Kraus operators and $A$ can be thus
chosen as the projection on $\mathcal{H}_S^1$ \footnote{If we use the
equivalence
$\mathcal{H}_1\otimes\mathcal{H}_2\otimes\mathcal{H}_1\otimes\mathcal{H}_2
\simeq
\mathcal{H}_1\otimes\mathcal{H}_1\otimes\mathcal{H}_2\otimes\mathcal{H}_2$,
the subspace $\mathcal{H}_S^1$ can be identified with
$\mathcal{H}_1\wedge\mathcal{H}_1\otimes\mathcal{H}_2\wedge\mathcal{H}_2$.}.
Similar analysis may be also performed in the case of bosons and fermions. In
both cases $A$ is a projector on subspaces $\mathcal{H}^B_-$ and
$\mathcal{H}^F_-$ described in details in the Appendix.

\section{Mixed states}

Every mixed state $\rho$ can be decomposed as a convex combination of pure
states
\begin{equation}\label{rho0}
\rho=\sum_{k=1}^Kp_k\kb{\psi_k}{\psi_k}, \quad p_k>0,
\end{equation}
which, with $\ket{\phi_k}=\sqrt{p_k}\ket\psi_k$, can be rewritten as
\begin{equation}\label{rho1}
\rho=\sum_{k=1}^K\kb{\phi_k}{\phi_k}.
\end{equation}
A particular example of (\ref{rho0}) is provided by the spectral
decomposition of $\rho$,
\begin{equation}\label{sprho0}
\rho=\sum_{k=1}^R r_k\kb{\eta_k}{\eta_k}, \quad \rho\ket{\eta_k}
=r_k\ket{\eta_k},\quad R=\mathrm{rank }\rho,
\end{equation}
or, equivalently, as in (\ref{rho1})
\begin{equation}\label{sprho1}
\rho=\sum_{k=1}^R\kb{\xi_k}{\xi_k}, \quad\ket{\xi_k}=\sqrt{r_k}\ket{\eta_k}.
\end{equation}
In fact all other decompositions (\ref{rho1}) can be obtained from
(\ref{sprho1}) \textit{via}
\begin{equation}\label{hou}
\ket{\phi_k}=\sum_{j=1}^RV_{kj}\ket{\xi_j},
\end{equation}
where $V$ is a $K\times R$ matrix fulfilling $V^\dagger V=I$
\cite{hughston93}.

We may now define for a particular decomposition (\ref{rho1})
\begin{equation}\label{conc-dec}
c_A\big(\{\phi_k\}\big)=\sum_{k=1}^Kc_A(\phi_k),
\end{equation}
where $c_A(\psi)$ is given by (\ref{psiApsi}). It is now obvious that if
minimizing $c_A\big(\{\xi_k\}\big)$ over all decompositions (\ref{rho1}) of
$\rho$ gives zero then $\rho$ is nonentangled since it has a decomposition
into a convex combination of nonentangled pure states. We obtain in this way
a well defined measure of entanglement \cite{mkb04,mckb05} for mixed states
\begin{equation}\label{carho}
 c_A(\rho)=\min\sum_kc_a(\phi_k),
\end{equation}
where the minimum is taken over all decompositions (\ref{rho0}). Using
(\ref{kb}) and (\ref{hou}) we obtain further
\begin{eqnarray}\label{carho-exp}
C_A(\rho)&=&\min\sum_k C_A(\phi_k)=\min\sum_k\bra{\phi_k\otimes\phi_k}A
\ket{\phi_k\otimes\phi_k}^{1/2}
\nonumber \\
&=&\min\sum_k\big(V_{ki}^\ast V_{kj}^\ast V_{kl} V_{km}
\bra{\xi_i\otimes\xi_j}A\ket{\xi_l\otimes\phi_m}\big)^{1/2}
\nonumber \\
&=&\min\sum_k\Big(V_{ki}^\ast V_{kj}^\ast V_{kl} V_{km}
\sum_\mu\bra{\xi_i}T_\mu\ket{\xi_j^\ast}\bra{\xi_l^\ast}
T_\mu^\dagger\ket{\xi_m}\Big)^{1/2}
\nonumber \\
&=&\min\sum_k\Big(\sum_\mu(V^\ast\tau_\mu V^\dagger)_{kk}
(V\tau_\mu^\ast V^T)_{kk}\Big)^{1/2}
=\min\sum_k\Big(\sum_\mu|(V^\ast\tau_\mu
V^\dagger)_{kk}|^2\Big)^{1/2},
\end{eqnarray}
where $\tau_\mu$ are $r\times r$ matrices,
\begin{equation}\label{tau}
(\tau_\mu)_{ij}=\bra{\xi_i}T_\mu\ket{\xi_j^\ast},
\end{equation}
constructed from easily calculated ingredients: the eigenvectors of $\rho$
and the Kraus operators of $A$ (or, equivalently, $L$), and the minimum is
taken over all $V$ fulfilling $V^\dagger V$.

We may further reduce the complexity of minimization \cite{mkb04} by
employing the Cauchy-Schwartz inequality
\begin{equation}\label{C-S}
\left(\sum_\mu x_\mu^2\right)^{1/2}\left(\sum_\mu y_\mu^2\right)^{1/2}
\ge\sum_\mu x_\mu y_\mu.
\end{equation}
With $x_\mu=|(V^\ast\tau_\mu V^\dagger)_{kk}|$ and $\sum_\mu y_mu=1$. We
obtain thus
\begin{equation}\label{lb0}
c_A(\rho)\ge\min\sum_k\sum_\mu y_\mu|(V^\ast\tau_\mu
V^\dagger)_{kk}|\ge\min\sum_k
\left|\left[V^*\left(\sum_\mu y_\mu \tau_\mu\right)V^\dagger\right]_{kk}\right|,
\end{equation}
where we used $\sum_\mu|z_\mu|\ge\left|\sum_\mu z_\mu\right|$. The
minimization over $V$ can be now performed explicitly \cite{uhlmann00}
giving,
\begin{equation}\label{uhl}
c_A(\rho)\ge\max\left\{0,\lambda_1-\sum_{j>1}\lambda_j\right\},
\end{equation}
where $\lambda_j^2$ are the singular values of the matrix
$\mathcal{T}=\sum_\mu y_\mu \tau_\mu$. The matrix $\mathcal{T}$ still depends
on the parameters $y_\mu$ which can be chosen in an arbitrary way under the
condition $\sum_\mu y_\mu^2=1$, leaving a large freedom to construct lower
bounds for $c_A(\rho)$.

\section{Summary and conclusions}

We have presented a method of discriminating pure nonentangled states for
multipartite systems. The method is universal - it applies, at least in
principle, to systems with arbitrary number of distinguishable as well as
undistinguishable particles. Let us point at some other advantages of the
proposed approach
\begin{itemize}
\item The defined measure of entanglement is expressed in terms of a
    Hermitian, albeit bilinear operator -  ``a nonlinear entanglement
    witness'', and as such is, in principle, a physically measurable
    quantity.
\item The method, based solely on representation theory, can be easily
    adapted to more complicated situations, e.g.\ systems consisting of
    mixtures of bosons and fermions.
\item Calculation of the generalized concurrence is made in an
    algorithmic way consisting of few steps: 1) identification of a
    relevant group of local transformation and its representation, 2)
    calculation of the Lichtenstein's operator $L$ given in terms of the
    operators of the Lie algebra of the group, 3) identification of
    relevant Kraus operators of the image of $L$ under the
    Jamio{\l}kowski isomorphism. To perform the last step one looks for
    the symmetric Kraus operators which are obtained from eigenvectors of
    $L$ not belonging to its largest eigenvalue. The latter can be
    explicitly calculated from data about the group and representation in
    question.
\end{itemize}
The pure state generalized concurrence constructed here can be used as a
basis for effective estimates of mixed-state entanglement. One example of
such estimate has been presented in the last section of the paper.

\section{Acknowledgments}
The work was supported by SFB/TR12 'Symmetries and Universality in Mesoscopic
Systems' program of the Deutsche Forschungsgemeischaft and Polish MNiSW grant
no.\ DFG-SFB/38/2007.

\section{Appendix}

Let us choose the following basis in $\mathfrak{sl}_N(\mathbb{C})$
\begin{eqnarray}
X_{ij}&=&\kb{i}{j},\quad i,j=1,\ldots,N, \label{slNcr} \\
H_{l}&=&\frac{1}{\sqrt{l(l+1)}}
\left(\sum_{k=1}^l\kb{k}{k}-l\kb{l+1}{l+1}\right)
=\sum a_{lk}\kb{k}{k}, \quad l=1,\ldots,N-1. \label{slNcc}
\end{eqnarray}
The normalization of the basis elements was chosen to have $\tr{H_i^2}=1=\tr
X_{ij}X_{ji}$. The positive roots correspond to $i<j$ in (\ref{slNcr}).

Short calculations show that
\begin{equation}\label{lemma1}
\sum_{i=1}^{N-1} a_{ik} a_{il}=\left\{\begin{array}{lrr}
-\frac{1}{N} & \mathrm{for } & k\ne l \\
1-\frac{1}{N} & \mathrm{for } &k=l
\end{array}\right.
\end{equation}

Let us start with distinguishable particles. For simplicity we assume that
$\mathcal{H}_1=\mathcal{H}_2=:\mathcal{H}$, $\dim{H}=N$. Remember that
despite this we consider distinguishable particles, so we represent
$SU(N)\times SU(N)$ on $\mathcal{H}_{comp}=\mathcal{H}_1\otimes\mathcal{H}_2
=\mathcal{H}\otimes\mathcal{H}$. The corresponding representation of the Lie
algebra $\mathfrak{sl}_N(\mathbb{C})\oplus\mathfrak{sl}_N(\mathbb{C})$ is
given by
\begin{equation}\label{pid}
\pi((A,0))=A\otimes I,\quad \pi((0,B))=I\otimes B,
\end{equation}
hence
\begin{equation}\label{C2d}
C_2=\sum_{\alpha>0}\Big((X_\alpha X_{-\alpha}+X_{-\alpha}X_\alpha)
\otimes I+I\otimes(X_\alpha X_{-\alpha}+X_{-\alpha}X_\alpha)
\Big)+\sum_{j=1}^{N-1}\Big(H_j^2\otimes I+I\otimes H_j^2\Big).
\end{equation}
Using the explicit representation (\ref{slNcr})-(\ref{slNcc}) we obtain
\begin{equation}\label{C2d1}
C_2=\left(1-\frac{1}{N^2}\right)I\otimes I.
\end{equation}

Let $\ket{ijkl}:=\ket{i}\otimes\ket{j}\otimes\ket{k}\otimes\ket{l}
\in\mathcal{H}_{comp}\otimes\mathcal{H}_{comp}=
\mathcal{H}\otimes\mathcal{H}\otimes\mathcal{H}\otimes\mathcal{H}$ and let us
introduce two operators $S_1\ket{ijkl}=\ket{kjil}$,
$S_2\ket{ijkl}=\ket{ilkj}$. They define three orthogonal subspaces of
$\mathcal{H}_{comp}\otimes\mathcal{H}_{comp}$,
\begin{eqnarray}
\mathcal{H}_A&:=&\big\{\ket{\psi}:S_1S_2\ket{\psi}=-\ket{\psi}\big\}
=\mathrm{span}\big\{\ket{ijkl}-\ket{klij}\big\}, \label{spanHA} \\
\mathcal{H}_S^1&:=&\big\{\ket{\psi}:S_1\ket{\psi}
=S_2\ket{\psi}=-\ket{\psi}\big\}
=\mathrm{span}\big\{\ket{ijkl}+\ket{klij}-\ket{kjil}-\ket{ilkj}\big\},
\label{spanHS1}\\
\mathcal{H}_S^2&:=&\big\{\ket{\psi}:S_1\ket{\psi}
=S_2\ket{\psi}=\ket{\psi}\big\}
=\mathrm{span}\big\{\ket{ijkl}+\ket{klij}+\ket{kjil}+\ket{ilkj}\big\},
\label{spanHS2}
\end{eqnarray}
with $\mathcal{H}_A\oplus\mathcal{H}_S^1\oplus\mathcal{H}_S^2
=\mathcal{H}_{comp}\otimes\mathcal{H}_{comp}$. Applying $L$ to the vectors
spanning the subspaces written explicitly in (\ref{spanHA})-(\ref{spanHS2})
and using (\ref{lemma1}) and (\ref{C2d1}), we find that $\mathcal{H}_A$,
$\mathcal{H}_S^1$ and $\mathcal{H}_S^2$ are eigenspaces of $L$ with the
eigenvalues, respectively,
\begin{equation}\label{Leigend}
\lambda_A=2-\frac{4}{N^2}, \quad \lambda_S^1=2-\frac{2}{N}-\frac{4}{N^2},
\quad \lambda_S^2=2+\frac{2}{N}-\frac{4}{N^2}.
\end{equation}
The largest eigenvalue $l_{max}$ equals to $\lambda_S^2$, the space
$\mathcal{H}_S^2$ is thus in the kernel of the operator $A$ and does not
contribute Kraus operators to the generalized concurrence $c_A$
(\ref{psiApsi}). The Kraus operators constructed from the vectors in
$\mathcal{H}_A$ take the form $T\simeq \kb{ij}{kl}-\kb{kl}{ij}$ [cf.\
(\ref{T})], where $\ket{ij}=\ket{i}\otimes\ket{j}$ etc. They are thus
antisymmetric and, as explained in Section~\ref{sec:measures}, are also
irrelevant to $c_A$. We are left with the only ingredients given by the
subspace $H_S^1$, i.e.\ the Kraus operators
$T_{ijkl}=\kb{ij}{kl}+\kb{kl}{ij}-\kb{kj}{il}-\kb{il}{kj}$. Finally thus, in
the definition (\ref{psiApsi}) we can take as $A$ the projection on $H_S^1$,
which is a subspace of $\mathcal{H}_{comp}\otimes\mathcal{H}_{comp}=
\mathcal{H}\otimes\mathcal{H}\otimes\mathcal{H}\otimes\mathcal{H}$ obtained
by antisymmetrizing the first with the third factor as well as the second
with the fourth ones. Calculations for $\mathcal{H}_1\ne\mathcal{H}_2$ are
only slightly more complicated and give the same result, i.e.\ $A$ as the
projection on the subspace of
$\mathcal{H}_1\otimes\mathcal{H}_2\otimes\mathcal{H}_1\otimes\mathcal{H}_2$
obtained by separate antisymmetrizations of $\mathcal{H}_1$ factors (i.e.\
the first and third ones) and $\mathcal{H}_2$ (the second and fourth factor),
which coincides with the results of \cite{mkb04,mckb05}.

In the case of two identical particles with $N$-dimensional single-particle
spaces $\mathcal{H}$ the relevant representation is that of $SU(N)$ on
$\mathcal{H}_{comp}=\mathcal{H}\wedge\mathcal{H}$ or
$\mathcal{H}_{comp}=\mathcal{H}\vee\mathcal{H}$. The corresponding
representation of $\mathfrak{sl}_N(\mathbb{C})$ is given by
\begin{equation}\label{pii}
\pi(A)=A\otimes I+I\otimes A.
\end{equation}
As for the case of distinguishable particles we start with the calculation of
the Casimir operator $C_2$. To shorten the considerations we may calculate
$C_2$ on $\mathcal{H}\otimes\mathcal{H}$ for the representation (\ref{pii}),
for which it is not proportional to the identity since the representation on
the full tensor product is not irreducible. However, a further reduction to
the irreducible invariant subspaces $\mathcal{H}\wedge\mathcal{H}$ and
$\mathcal{H}\vee\mathcal{H}$ will lead to the desired results. Again from
(\ref{casimir}) and (\ref{pii}) we have
\begin{eqnarray}\label{C2di1}
C_2&=&\sum_{\alpha>0}\big(X_\alpha X_{-\alpha}\otimes I
+I\otimes X_\alpha X_{-\alpha}+2X_\alpha\otimes X_{-\alpha}
+X_{-\alpha} X_{\alpha}\otimes I
+I\otimes X_{-\alpha} X_{\alpha}+2X_{-\alpha}\otimes X_{\alpha}\big)
\nonumber \\
&+&\sum_{i=1}^{N-1}\big(H_i^2\otimes I+I\otimes H_i^2+
2H_i\otimes H_i).
\end{eqnarray}
Using the explicit form of $X_{ij}$ and $H_i$ given by (\ref{slNcr})and
(\ref{slNcc}) with the help of (\ref{lemma1}) we obtain
\begin{equation}\label{C2di2}
C_2=\left(1-\frac{2}{N^2}\pm\frac{1}{N}\right)I\otimes I.
\end{equation}
where the upper sign is for the symmetric subspace,
$\mathcal{H}\vee\mathcal{H}$, and the lower one for the antisymmetric one,
$\mathcal{H}\wedge\mathcal{H}$. As it should, in each of these subspaces the
Casimir operator is proportional to the identity, and in order to simplify
calculations it is convenient to consider the operator
\begin{eqnarray}\label{Lprime}
L^\prime&=&N\big(L-C_2\otimes I-I\otimes C_2\big) \nonumber \\
&=&2N\sum_{\alpha>0}\Big(\pi(X_\alpha)\otimes\pi(X_{-\alpha})
+\pi(X_{-\alpha})\otimes\pi(X_{\alpha})\Big)+
2N\sum_{i=1}^r \pi(H_i)\otimes\pi(H_i).
\end{eqnarray}
Substituting (\ref{slNcr})and (\ref{slNcc}) we obtain after straightforward
calculations
\begin{eqnarray}\label{Lprimeaction}
L^\prime\ket{ijkl}&=&(1-\delta_{ik})\ket{kjil}+(1-\delta_{il})\ket{ljki}+
(1-\delta_{jk})\ket{ikjl}+(1-\delta_{jl})\ket{ilkj}
\nonumber \\
&+&\left(-\frac{4}{N}+\delta_{ik}+\delta_{il}+\delta_{jk}+\delta_{jl}\right)
\ket{ijkl},
\end{eqnarray}
for
$\ket{ijkl}\in\mathcal{H}\otimes\mathcal{H}\otimes\mathcal{H}\otimes\mathcal{H}$.

We are now in the position where we have to specify further calculations to
$\mathcal{H}_{comp}=\mathcal{H}\vee\mathcal{H}$ (bosons) or
$\mathcal{H}_{comp}=\mathcal{H}\wedge\mathcal{H}$ (fermions).

\subsection{Bosons}
We choose a basis in $\mathcal{H}_{comp}=\mathcal{H}\vee\mathcal{H}$
consisting of vectors
\begin{eqnarray}\label{bbasis}
\ket{\psi_i}&=&\ket{i}\otimes\ket{i}, \quad j=1,\ldots,N \\
\ket{\psi_{ij}}&=&\ket{i}\otimes\ket{j}+\ket{j}\otimes\ket{i},
\quad I,j=1,\ldots,N, \quad i\ne j,
\end{eqnarray}
and split the space $\mathcal{H}_{comp}\otimes\mathcal{H}_{comp}=
\mathcal{H}\vee\mathcal{H}\otimes\mathcal{H}\vee\mathcal{H}$ into two parts,
$\mathcal{H}_S^B$ which is symmetric with respect to interchange of the two
copies of $\mathcal{H}_{comp}$ and $\mathcal{H}_A^B$ which is antisymmetric
with respect to this interchange. The space $\mathcal{H}_S^B$ can be
decomposed into invariant spaces of the operator $L^\prime$,
\begin{enumerate}
\item $\mathcal{H}_{1}^B$ spanned by vectors
    $\ket{\psi_{i}}\otimes\ket{\psi_{i}}$,
\item $\mathcal{H}_{2}^B$ spanned by vectors
    $\ket{\psi_{i}}\otimes\ket{\psi_{j}}+
    \ket{\psi_{j}}\otimes\ket{\psi_{i}}$ and
    $\ket{\psi_{ij}}\otimes\ket{\psi_{ij}}$, $i\ne j$,
\item $\mathcal{H}_{3}^B$ spanned by vectors
    $\ket{\psi_{i}}\otimes\ket{\psi_{ij}}+
    \ket{\psi_{ij}}\otimes\ket{\psi_{i}}$, $i\ne j$,
\item $\mathcal{H}_{4}^B$ spanned by vectors
    $\ket{\psi_{i}}\otimes\ket{\psi_{jk}}+
    \ket{\psi_{jk}}\otimes\ket{\psi_{i}}$ and
    $\ket{\psi_{ij}}\otimes\ket{\psi_{ik}}+
    \ket{\psi_{ik}}\otimes\ket{\psi_{ij}}$, $i\ne j\ne k\ne i$,
\item $\mathcal{H}_{5}^B$ spanned by vectors
    $\ket{\psi_{ij}}\otimes\ket{\psi_{kl}}+
    \ket{\psi_{kl}}\otimes\ket{\psi_{ij}}$,
    $\ket{\psi_{ik}}\otimes\ket{\psi_{jl}}+
    \ket{\psi_{jl}}\otimes\ket{\psi_{ik}}$, and
    $\ket{\psi_{il}}\otimes\ket{\psi_{jk}}+
    \ket{\psi_{jk}}\otimes\ket{\psi_{il}}$ with all $i,j,k,l$ different.
\end{enumerate}
Using (\ref{Lprimeaction}) we can diagonalize $L^\prime$ in each of the above
subspaces. In particular we have:
\begin{enumerate}

\item In the subspace $\mathcal{H}_{1}^B$
\begin{equation}\label{B1}
L^\prime\ket{\psi_{i}}\otimes\ket{\psi_{i}}=
\left(4-\frac{4}{N}\right)\ket{\psi_{i}}\otimes\ket{\psi_{i}}
=\lambda_+\ket{\psi_{i}}\otimes\ket{\psi_{i}}.
\end{equation}

\item In the subspace $\mathcal{H}_{2}^B$
\begin{eqnarray}\label{B2}
&&
L^\prime\big(\ket{\psi_{i}}\otimes\ket{\psi_{j}}+
    \ket{\psi_{j}}\otimes\ket{\psi_{i}}\big)=
-\frac{4}{N}\big(\ket{\psi_{i}}\otimes\ket{\psi_{j}}+
    \ket{\psi_{j}}\otimes\ket{\psi_{i}}\big)
    +2\ket{\psi_{ij}}\otimes\ket{\psi_{ij}} \\
&&
L^\prime\ket{\psi_{ij}}\otimes\ket{\psi_{ij}}=
4\big(\ket{\psi_{i}}\otimes\ket{\psi_{j}}+\ket{\psi_{j}}\otimes\ket{\psi_{i}}\big)
+\left(2-\frac{4}{N}\right)\ket{\psi_{ij}}\otimes\ket{\psi_{ij}}.
\end{eqnarray}
A consequent diagonalization of the $2\times 2$ matrix,
\begin{equation}\label{B2M}
\left[
\begin{array}{cc}
 -\frac{4}{N} &       4       \\
      2       & 2-\frac{4}{N} \\
\end{array}
\right],
\end{equation}
reveals that
\begin{eqnarray}
L^\prime\ket{\Psi_{ij}^1}&=\left(4-\frac{4}{N}\right)\ket{\Psi_{ij}^1}=
\lambda_+\ket{\Psi_{ij}^1}, \label{B2final1} \\
L^\prime\ket{\Psi_{ij}^2}&=\left(-2-\frac{4}{N}\right)\ket{\Psi_{ij}^2}=
\lambda_-\ket{\Psi_{ij}^2}, \label{B2final2}
\end{eqnarray}
where
\begin{eqnarray}
\ket{\Psi_{ij}^1}&=&\ket{\psi_{i}}\otimes\ket{\psi_{j}}+
\ket{\psi_{j}}\otimes\ket{\psi_{i}}
+\ket{\psi_{ij}}\otimes\ket{\psi_{ij}} \label{B2final1a} \\
\ket{\Psi_{ij}^2}&=&-2\ket{\psi_{i}}\otimes\ket{\psi_{j}}
-2\ket{\psi_{j}}\otimes\ket{\psi_{i}}
+\ket{\psi_{ij}}\otimes\ket{\psi_{ij}}. \label{B2final2a}
\end{eqnarray}

\item In the subspace $\mathcal{H}_{3}^B$
\begin{eqnarray}\label{B3}
L^\prime\big(\ket{\psi_{i}}\otimes\ket{\psi_{ij}}+
\ket{\psi_{ij}}\otimes\ket{\psi_{i}} \big)
&=&\left(4-\frac{4}{N}\right)\big(\ket{\psi_{i}}\otimes\ket{\psi_{ij}}+
\ket{\psi_{ij}}\otimes\ket{\psi_{i}} \big) \nonumber \\
&=&\lambda_+\big(\ket{\psi_{i}}\otimes\ket{\psi_{ij}}+
\ket{\psi_{ij}}\otimes\ket{\psi_{i}} \big)
\end{eqnarray}

\item In the subspace $\mathcal{H}_{4}^B$
\begin{eqnarray}\label{B4}
L^\prime\big(\ket{\psi_{i}}\otimes\ket{\psi_{jk}}+
    \ket{\psi_{jk}}\otimes\ket{\psi_{i}}\big)&=&
\left(-\frac{4}{N}\right)\big(\ket{\psi_{i}}\otimes\ket{\psi_{jk}}+
\ket{\psi_{jk}}\otimes\ket{\psi_{i}}\big)
\nonumber \\
&+&2\big(\ket{\psi_{ij}}\otimes\ket{\psi_{ik}}+
    \ket{\psi_{ik}}\otimes\ket{\psi_{ij}}\big) \\
L^\prime\big(\ket{\psi_{ij}}\otimes\ket{\psi_{ik}}+
    \ket{\psi_{ik}}\otimes\ket{\psi_{ij}}\big)&=&
4\big(\ket{\psi_{i}}\otimes\ket{\psi_{jk}}+
\ket{\psi_{jk}}\otimes\ket{\psi_{i}}\big)
\nonumber \\
&+&\left(2-\frac{4}{N}\right)\big(\ket{\psi_{ij}}\otimes\ket{\psi_{ik}}+
    \ket{\psi_{ik}}\otimes\ket{\psi_{ij}}\big).
\end{eqnarray}
Hence, like in 2. above,
\begin{eqnarray}
L^\prime\ket{\Psi_{ijk}^3}&=\left(4-\frac{4}{N}\right)\ket{\Psi_{ijk}^3}=
\lambda_+\ket{\Psi_{ijk}^3}, \label{B4final1}\\
L^\prime\ket{\Psi_{ijk}^4}&=\left(-2-\frac{4}{N}\right)\ket{\Psi_{ijk}^4}=
\lambda_-\ket{\Psi_{ijk}^4}, \label{B4final2}
\end{eqnarray}
where
\begin{eqnarray}
\ket{\Psi_{ijk}^3}&=&\ket{\psi_{i}}\otimes\ket{\psi_{jk}}+
\ket{\psi_{jk}}\otimes\ket{\psi_{i}}
+\ket{\psi_{ij}}\otimes\ket{\psi_{ik}}+\ket{\psi_{ik}}\otimes\ket{\psi_{ij}}
\label{B4final1a} \\
\ket{\Psi_{ijk}^4}&=&-2\ket{\psi_{i}}\otimes\ket{\psi_{jk}}-
2\ket{\psi_{jk}}\otimes\ket{\psi_{i}}
+\ket{\psi_{ij}}\otimes\ket{\psi_{ik}}+\ket{\psi_{ik}}\otimes\ket{\psi_{ij}}.
\label{B4final2a}
\end{eqnarray}

\item In the subspace $\mathcal{H}_{5}^B$
\begin{eqnarray}\label{B4a}
L^\prime\big(\ket{\psi_{ij}}\otimes\ket{\psi_{kl}}+
    \ket{\psi_{kl}}\otimes\ket{\psi_{ij}}\big)&=&
-\frac{4}{N}\big(\ket{\psi_{kl}}\otimes\ket{\psi_{ij}}+
\ket{\psi_{ij}}\otimes\ket{\psi_{kl}}\big)
\nonumber \\
&+&2\big(\ket{\psi_{ik}}\otimes\ket{\psi_{jl}}+
    \ket{\psi_{jl}}\otimes\ket{\psi_{ik}}\big)
\nonumber \\
&+&2\big(\ket{\psi_{il}}\otimes\ket{\psi_{jk}}+
    \ket{\psi_{jk}}\otimes\ket{\psi_{il}}\big)
\\
L^\prime\big(\ket{\psi_{ik}}\otimes\ket{\psi_{jl}}+
    \ket{\psi_{jl}}\otimes\ket{\psi_{ik}}\big)&=&
2\big(\ket{\psi_{kl}}\otimes\ket{\psi_{ij}}+
\ket{\psi_{ij}}\otimes\ket{\psi_{kl}}\big)
\nonumber \\
&-&\frac{4}{N}\big(\ket{\psi_{ik}}\otimes\ket{\psi_{jl}}+
    \ket{\psi_{jl}}\otimes\ket{\psi_{ik}}\big)
\nonumber \\
&+&2\big(\ket{\psi_{il}}\otimes\ket{\psi_{jk}}+
    \ket{\psi_{jk}}\otimes\ket{\psi_{il}}\big)
\\
L^\prime\big(\ket{\psi_{il}}\otimes\ket{\psi_{jk}}+
    \ket{\psi_{jk}}\otimes\ket{\psi_{il}}\big)&=&
2\big(\ket{\psi_{kl}}\otimes\ket{\psi_{ij}}+
\ket{\psi_{ij}}\otimes\ket{\psi_{kl}}\big)
\nonumber \\
&+&2\big(\ket{\psi_{ik}}\otimes\ket{\psi_{jl}}+
    \ket{\psi_{jl}}\otimes\ket{\psi_{ik}}\big)
\nonumber \\
&-&\frac{4}{N}\big(\ket{\psi_{il}}\otimes\ket{\psi_{jk}}+
    \ket{\psi_{jk}}\otimes\ket{\psi_{il}}\big).
\end{eqnarray}
The relevant matrix to diagonalize reads as
\begin{equation}\label{B4Ma}
\left[
\begin{array}{ccc}
 -\frac{4}{N} & 2 & 2      \\
 2 & -\frac{4}{N} & 2      \\
 2 & 2 & -\frac{4}{N}      \\
\end{array}
\right],
\end{equation}
which leads to
\begin{eqnarray}\label{B4finala}
L^\prime\ket{\Psi_{ijkl}^1}&=\left(4-\frac{4}{N}\right)\ket{\Psi_{ijkl}^1}=
\lambda_+\ket{\Psi_{ijkl}^1}, \label{B4afinal1}\\
L^\prime\ket{\Psi_{ijkl}^2}&=\left(-2-\frac{4}{N}\right)\ket{\Psi_{ijkl}^2}=
\lambda_-\ket{\Psi_{ijkl}^2}, \label{B4afinal2} \\
L^\prime\ket{\Psi_{ijkl}^3}&=\left(-2-\frac{4}{N}\right)\ket{\Psi_{ijkl}^3}=
\lambda_-\ket{\Psi_{ijkl}^3}, \label{B4afinal3}
\end{eqnarray}
with
\begin{eqnarray}
\ket{\Psi_{ijkl}^1}
&=&\ket{\psi_{ij}}\otimes\ket{\psi_{kl}}+
\ket{\psi_{kl}}\otimes\ket{\psi_{ij}}+
\ket{\psi_{ik}}\otimes\ket{\psi_{jl}}
\nonumber  \\
&+&\ket{\psi_{jl}}\otimes\ket{\psi_{ik}}
+\ket{\psi_{il}}\otimes\ket{\psi_{jk}}
+\ket{\psi_{jk}}\otimes\ket{\psi_{il}}
\label{B4afinal1a} \\
\ket{\Psi_{ijkl}^2}
&=&\ket{\psi_{ij}}\otimes\ket{\psi_{kl}}+
\ket{\psi_{kl}}\otimes\ket{\psi_{ij}}
-\ket{\psi_{il}}\otimes\ket{\psi_{jk}}
-\ket{\psi_{jk}}\otimes\ket{\psi_{il}},
\label{B4afinal2a} \\
\ket{\Psi_{ijkl}^3}
&=&\ket{\psi_{ij}}\otimes\ket{\psi_{kl}}+
\ket{\psi_{kl}}\otimes\ket{\psi_{ij}}
-\ket{\psi_{ik}}\otimes\ket{\psi_{jl}}
-\ket{\psi_{jl}}\otimes\ket{\psi_{ik}}.
\label{B4afinal3a}
\end{eqnarray}
\end{enumerate}

To finish the calculations in the bosonic case we have to consider the
antisymmetric part
$\mathcal{H}_A^B=\mathcal{H}\vee\mathcal{H}\wedge\mathcal{H}\vee\mathcal{H}$
of the space $\mathcal{H}_{comp}\otimes\mathcal{H}_{comp}$. It decomposes
into invariant subspaces
\begin{enumerate}
\item $\mathcal{H}_6^B$ spanned by vectors
    $\ket{\tilde\Psi_{ij}^1}=\ket{\psi_{i}}\otimes\ket{\psi_{j}}-
    \ket{\psi_{j}}\otimes\ket{\psi_{i}}$, $i\ne j$,
\item $\mathcal{H}_7^B$ spanned by vectors
    $\ket{\tilde\Psi_{ij}^2}=\ket{\psi_{i}}\otimes\ket{\psi_{ij}}-
    \ket{\psi_{ij}}\otimes\ket{\psi_{i}}$, $i\ne j$,
\item $\mathcal{H}_8^B$ spanned by vectors
    $\ket{\tilde\Psi_{ijk}^1}=\ket{\psi_{i}}\otimes\ket{\psi_{jk}}-
    \ket{\psi_{jk}}\otimes\ket{\psi_{i}}$ and
    $\ket{\tilde\Psi_{ijk}^2}=\ket{\psi_{ij}}\otimes\ket{\psi_{ik}}-
    \ket{\psi_{ik}}\otimes\ket{\psi_{ij}}$, $i\ne j\ne k$,
\item $\mathcal{H}_9^B$ spanned by vectors
    $\ket{\tilde\Psi_{ijkl}^1}=\ket{\psi_{ij}}\otimes\ket{\psi_{kl}}-
    \ket{\psi_{kl}}\otimes\ket{\psi_{ij}}$,
    $\ket{\tilde\Psi_{ijkl}^2}=\ket{\psi_{ik}}\otimes\ket{\psi_{jl}}-
    \ket{\psi_{jl}}\otimes\ket{\psi_{ik}}$, and
    $\ket{\tilde\Psi_{ijkl}^3}=\ket{\psi_{il}}\otimes\ket{\psi_{jk}}-
    \ket{\psi_{jk}}\otimes\ket{\psi_{il}}$ with all $i,j,k,l$ different.
\end{enumerate}
From (\ref{Lprimeaction}) we obtain straightforwardly
\begin{equation}\label{BA}
L^\prime\ket{\tilde\Psi}=-\frac{4}{N}\ket{\tilde\Psi}
=\lambda\ket{\tilde\Psi}
\end{equation}
for all the subspaces, i.e. for
$\ket{\tilde\Psi}=\ket{\tilde\Psi_{ij}^{1,2}},\ket{\tilde\Psi_{ijk}^{1,2}}$,
and $\ket{\tilde\Psi_{ijkl}^{1,2,3}}$.

Furthermore, we easily see that if we do not assume that all $i, j, k, l$ are
different in the definitions of $\ket{\Psi_{ijkl}^{1, 2, 3}}$ and
$\ket{\tilde\Psi_{ijkl}^{1, 2, 3}}$ (adopting the notation $\ket{\psi_{ii}} =
2\ket{ii}$), we can express all vectors spanning the subspaces
$\mathcal{H}_i^B$, $i = 1, \ldots, 9$, in terms of $\ket{\Psi_{ijkl}^{1, 2,
3}}$ and $\ket{\tilde\Psi_{ijkl}^{1}}$ (for example, $\ket{\psi_i}\otimes
\ket{\psi_{ij}}+\ket{\psi_{ij}}\otimes\ket{\psi_i} =
\frac{1}{6}\ket{\Psi_{iiij}^{1}}$ etc.).

To sum it up, the Hilbert space $\mathcal{H}_{comp}\otimes\mathcal{H}_{comp}=
\mathcal{H}\vee\mathcal{H}\otimes\mathcal{H}\vee\mathcal{H}$ splits into
three eigenspaces of $L^\prime$,
\begin{equation}\label{splitbosonplus}
\mathcal{H}^B_+=\mathrm{span}\big\{\ket{\Psi_{ijkl}^1}\big\},
\end{equation}
[see Equations
(\ref{B1}),(\ref{B2final1}),(\ref{B2final1a}),(\ref{B3}),(\ref{B4final1}),(\ref{B4final1a}),(\ref{B4afinal1}),
and (\ref{B4afinal1a})],
\begin{equation}\label{splitbosonzero}
\mathcal{H}^B=\mathrm{span}\big\{\ket{\tilde\Psi_{ijkl}^1}, \ket{\tilde\Psi_{ijkl}^2}, \ket{\tilde\Psi_{ijkl}^3}\big\},
\end{equation}
[see Equation (\ref{BA})], and
\begin{equation}\label{splitbosonminus}
\mathcal{H}^B_-=\mathrm{span}\big\{\ket{\Psi_{ijkl}^2}, \ket{\Psi_{ijkl}^3} \big\},
\end{equation}
for all $i, j, k, l$, not necessarily different - see the remark above, [see
Equations
(\ref{B2final2}),(\ref{B2final2a}),(\ref{B4final2}),(\ref{B4final2a}),(\ref{B4afinal2}),(\ref{B4afinal3}),(\ref{B4afinal2a}),
and(\ref{B4afinal3a})].

They correspond to three different eigenvalues of $L^\prime$, respectively,
$\lambda_+=4-\frac{4}{N}$, $\lambda=-\frac{4}{N}$, and
$\lambda_+=-2-\frac{4}{N}$. The subspaces
(\ref{splitbosonplus}-\ref{splitbosonminus}) are also eigenspaces of
$L=\frac{1}{N}L^\prime+2\left(1-\frac{2}{N^2}+\frac{1}{N}\right) I$, and the
corresponding eigenvalues of $L$ are
$\lambda^L_+=2-\frac{8}{N^2}+\frac{6}{N}$,
$\lambda^L=2-\frac{8}{N^2}+\frac{2}{N}$, and $\lambda^L_-=2-\frac{8}{N^2}$.
The largest one, $l_{max}=\lambda^L_+$ corresponds to the kernel of the
operator $A=l_{max}I-L$, and vectors in the subspace $\mathcal{H}^B$ give
antisymmetric Kraus operators $T_\mu$. Hence as $A$ we may take the
projection on $\mathcal{H}^B_-$ and construct the relevant Kraus operators
from vectors in this space.

\subsection{Fermions}

Calculations for fermions closely follow the bosonic case, so we omit most of
details. In the space $\mathcal{H}_{comp}=\mathcal{H}\wedge\mathcal{H}$ we
choose a basis
$\big\{\ket{\phi_{ij}}=\ket{i}\otimes\ket{j}-\ket{j}\otimes\ket{i},i\ne
j\big\}$. As in the bosonic case we decompose
$\mathcal{H}_{comp}\otimes\mathcal{H}_{comp}$ into the symmetric
$\mathcal{H}_S^F=\mathcal{H}\wedge\mathcal{H}\vee\mathcal{H}\wedge\mathcal{H}$,
and antisymmetric,
$\mathcal{H}_A^F=\mathcal{H}\wedge\mathcal{H}\wedge\mathcal{H}\wedge\mathcal{H}$,
parts. The symmetric part can be further split into invariant subspaces of
$L^\prime$
\begin{enumerate}

\item $\mathcal{H}_{1}^F$ spanned by vectors
    $\ket{\Phi_{ij}^1}=\ket{\phi_{ij}}\otimes\ket{\phi_{ij}}$ for which
\begin{equation}\label{FS1}
L^\prime\ket{\Phi_{ij}^1}=\left(2-\frac{4}{N}\right)\ket{\Phi_{ij}^1}
=\lambda^\prime_+\ket{\Phi_{ij}^1}
\end{equation}

\item $\mathcal{H}_{2}^F$ spanned by vectors
    $\ket{\Phi_{ij}^2}=\ket{\phi_{ij}}\otimes\ket{\phi_{ik}}+
    \ket{\phi_{ik}}\otimes\ket{\phi_{ij}}$, $ j\ne k$ for which again
\begin{equation}\label{FS2}
L^\prime\ket{\Phi_{ij}^2}=\left(2-\frac{4}{N}\right)\ket{\Phi_{ij}^2}
=\lambda^\prime_+\ket{\Phi_{ij}^2}
\end{equation}

\item $\mathcal{H}_{3}^F$ spanned by vectors
    $\ket{\phi_{ij}}\otimes\ket{\phi_{kl}}+
    \ket{\phi_{kl}}\otimes\ket{\phi_{ij}}$,
    $\ket{\phi_{ik}}\otimes\ket{\phi_{jl}}+
    \ket{\phi_{jl}}\otimes\ket{\phi_{ik}}$, and
    $\ket{\phi_{il}}\otimes\ket{\phi_{jk}}+
    \ket{\phi_{jk}}\otimes\ket{\phi_{il}}$ with all $i,j,k,l$ different.

Here we diagonalize a set of equations
\begin{eqnarray}\label{FS3}
L^\prime\big(\ket{\phi_{ij}}\otimes\ket{\phi_{kl}}+
    \ket{\phi_{kl}}\otimes\ket{\phi_{ij}}\big)&=&
-\frac{4}{N}\big(\ket{\phi_{kl}}\otimes\ket{\phi_{ij}}+
\ket{\phi_{ij}}\otimes\ket{\phi_{kl}}\big)
\nonumber \\
&+&2\big(\ket{\phi_{ik}}\otimes\ket{\phi_{jl}}+
    \ket{\phi_{jl}}\otimes\ket{\phi_{ik}}\big)
\nonumber \\
&+&2\big(\ket{\phi_{il}}\otimes\ket{\phi_{jk}}+
    \ket{\phi_{jk}}\otimes\ket{\phi_{il}}\big)
\\
L^\prime\big(\ket{\phi_{ik}}\otimes\ket{\phi_{jl}}+
    \ket{\phi_{jl}}\otimes\ket{\phi_{ik}}\big)&=&
2\big(\ket{\phi_{kl}}\otimes\ket{\phi_{ij}}+
\ket{\phi_{ij}}\otimes\ket{\phi_{kl}}\big)
\nonumber \\
&-&\frac{4}{N}\big(\ket{\phi_{ik}}\otimes\ket{\phi_{jl}}+
    \ket{\phi_{jl}}\otimes\ket{\phi_{ik}}\big)
\nonumber \\
&-&2\big(\ket{\phi_{il}}\otimes\ket{\phi_{jk}}+
    \ket{\phi_{jk}}\otimes\ket{\phi_{il}}\big)
\\
L^\prime\big(\ket{\phi_{il}}\otimes\ket{\phi_{jk}}+
    \ket{\phi_{jk}}\otimes\ket{\phi_{il}}\big)&=&
2\big(\ket{\phi_{kl}}\otimes\ket{\phi_{ij}}+
\ket{\phi_{ij}}\otimes\ket{\phi_{kl}}\big)
\nonumber \\
&-&2\big(\ket{\phi_{ik}}\otimes\ket{\phi_{jl}}+
    \ket{\phi_{jl}}\otimes\ket{\phi_{ik}}\big)
\nonumber \\
&-&\frac{4}{N}\big(\ket{\phi_{il}}\otimes\ket{\phi_{jk}}+
    \ket{\phi_{jk}}\otimes\ket{\phi_{il}}\big).
\end{eqnarray}
producing
\begin{eqnarray}\label{FS3final}
L^\prime\ket{\Phi_{ijkl}^1}&=&\left(-4-\frac{4}{N}\right)\ket{\Phi_{ijkl}^1}=
\lambda^\prime_-\ket{\Phi_{ijkl}^1}, \\
L^\prime\ket{\Phi_{ijkl}^2}&=&\left(2-\frac{4}{N}\right)\ket{\Phi_{ijkl}^2}=
\lambda^\prime_+\ket{\Phi_{ijkl}^2}, \\
L^\prime\ket{\Phi_{ijkl}^3}&=&\left(2-\frac{4}{N}\right)\ket{\Phi_{ijkl}^3}=
\lambda^\prime_+\ket{\Phi_{ijkl}^3},
\end{eqnarray}
with
\begin{eqnarray}
\ket{\Phi_{ijkl}^1}
&=&-\ket{\phi_{ij}}\otimes\ket{\phi_{kl}}-
\ket{\phi_{kl}}\otimes\ket{\phi_{ij}}+
\ket{\phi_{ik}}\otimes\ket{\phi_{jl}}
\nonumber \\
&+&\ket{\phi_{jl}}\otimes\ket{\phi_{ik}}
+\ket{\phi_{il}}\otimes\ket{\phi_{jk}}
+\ket{\phi_{jk}}\otimes\ket{\phi_{il}}
\\
\ket{\Phi_{ijkl}^2}
&=&\ket{\phi_{ij}}\otimes\ket{\phi_{kl}}+
\ket{\phi_{kl}}\otimes\ket{\phi_{ij}}
+\ket{\phi_{il}}\otimes\ket{\phi_{jk}}
+\ket{\phi_{jk}}\otimes\ket{\phi_{il}},
\\
\ket{\Phi_{ijkl}^3}
&=&\ket{\phi_{ij}}\otimes\ket{\phi_{kl}}+
\ket{\phi_{kl}}\otimes\ket{\phi_{ij}}
+\ket{\phi_{ik}}\otimes\ket{\phi_{jl}}
+\ket{\phi_{jl}}\otimes\ket{\phi_{ik}}.
\end{eqnarray}
\end{enumerate}

The antisymmetric part $\mathcal{H}_A^F$ splits into invariant subspaces
\begin{enumerate}
\item $\mathcal{H}_4^F$ spanned by vectors
    $\ket{\tilde\Phi_{ijk}^1}=\ket{\phi_{ij}}\otimes\ket{\phi_{ik}}-
    \ket{\phi_{ik}}\otimes\ket{\phi_{ij}}$, $j\ne k$,
\item $\mathcal{H}_5^F$ spanned by vectors
    $\ket{\tilde\Phi_{ijkl}^1}=\ket{\phi_{ij}}\otimes\ket{\phi_{kl}}-
    \ket{\phi_{kl}}\otimes\ket{\phi_{ij}}$,
    $\ket{\tilde\Phi_{ijkl}^2}=\ket{\phi_{ik}}\otimes\ket{\phi_{jl}}-
    \ket{\phi_{jl}}\otimes\ket{\phi_{ik}}$, and
    $\ket{\tilde\Phi_{ijkl}^3}=\ket{\phi_{il}}\otimes\ket{\phi_{jk}}-
    \ket{\phi_{jk}}\otimes\ket{\phi_{il}}$ with all $i,j,k,l$ different.
\end{enumerate}
In both subspaces $L^\prime$ acts as the multiplication by
$\lambda^\prime=-\frac{4}{N}$.

As in the bosonic case, we can express all vectors spanning subspaces
$\mathcal{H}_i^F$, $i=1, \ldots, 5$, using only $\ket{\Phi_{ijkl}^{1, 2, 3}}$
and $\ket{\tilde\Phi_{ijkl}^{1, 2, 3}}$ (here we have $\ket{\phi_{ii}} = 0$).

Therefore $L^\prime$ has three different eigenvalues $\lambda^\prime_+$,
$\lambda^\prime$, and $\lambda^\prime_-$ corresponding to three eigenspaces,
\begin{equation}\label{splitfermionplus}
\mathcal{H}^F_+=\mathrm{span}\big\{\ket{\Phi_{ijkl}^2},\ket{\Phi_{ijkl}^3}\big\},
\end{equation}
\begin{equation}\label{splitfermionzero}
\mathcal{H}^F=\mathrm{span}\big\{\ket{\tilde\Phi_{ijkl}^1}, \ket{\tilde\Phi_{ijkl}^2}, \ket{\tilde\Phi_{ijkl}^3}\big\},
\end{equation}
and
\begin{equation}\label{splitfermionminus}
\mathcal{H}^F_-=\mathrm{span}\big\{\ket{\Phi_{ijkl}^1}\big\},
\end{equation}
for all $i, j, k, l$, not necessarily different. These are also the
eigenspaces of
$L=\frac{1}{N}L^\prime+2\left(1-\frac{2}{N^2}-\frac{1}{N}\right) I$. The
subspace $\mathcal{H}^F_+$ corresponding to the largest eigenvalue of $L$,
$l_{max}=\lambda^L_+=2-\frac{8}{N^2}$, is in the kernel of $A$, whereas
$\mathcal{H}^F$ corresponding to the middle eigenvalue
$\lambda^L=2-\frac{8}{N^2}-\frac{2}{N}$ of $L$ gives rise to antisymmetric
Kraus operators. The operator $A$ can be thus chosen as the projection on
$\mathcal{H}^F_-$.

%\bibliographystyle{apsrev}
%\bibliography{classtates}

\begin{thebibliography}{32}
\expandafter\ifx\csname natexlab\endcsname\relax\def\natexlab#1{#1}\fi
\expandafter\ifx\csname bibnamefont\endcsname\relax
  \def\bibnamefont#1{#1}\fi
\expandafter\ifx\csname bibfnamefont\endcsname\relax
  \def\bibfnamefont#1{#1}\fi
\expandafter\ifx\csname citenamefont\endcsname\relax
  \def\citenamefont#1{#1}\fi
\expandafter\ifx\csname url\endcsname\relax
  \def\url#1{\texttt{#1}}\fi
\expandafter\ifx\csname urlprefix\endcsname\relax\def\urlprefix{URL }\fi
\providecommand{\bibinfo}[2]{#2} \providecommand{\eprint}[2][]{\url{#2}}

\bibitem[{\citenamefont{Bengtsson and \.Zyczkowski}(2006)}]{bengtsson06}
    \bibinfo{author}{\bibfnamefont{I.}~\bibnamefont{Bengtsson}}
    \bibnamefont{and}
  \bibinfo{author}{\bibfnamefont{K.}~\bibnamefont{\.Zyczkowski}},
  \emph{\bibinfo{title}{{G}eometry of {Q}uantum {S}tates}}
  (\bibinfo{publisher}{Cambridge University Press},
  \bibinfo{address}{Cambridge}, \bibinfo{year}{2006}).

\bibitem[{\citenamefont{Badzi\c{a}g et~al.}(2002)\citenamefont{Badzi\c{a}g,
  Deuar, Horodecki, Horodecki, and Horodecki}}]{badziag02}
\bibinfo{author}{\bibfnamefont{P.}~\bibnamefont{Badzi\c{a}g}},
  \bibinfo{author}{\bibfnamefont{P.}~\bibnamefont{Deuar}},
  \bibinfo{author}{\bibfnamefont{M.}~\bibnamefont{Horodecki}},
  \bibinfo{author}{\bibfnamefont{P.}~\bibnamefont{Horodecki}},
  \bibnamefont{and}
  \bibinfo{author}{\bibfnamefont{R.}~\bibnamefont{Horodecki}},
  \bibinfo{journal}{J. Mod. Opt.} \textbf{\bibinfo{volume}{49}},
  \bibinfo{pages}{1289} (\bibinfo{year}{2002}).

\bibitem[{\citenamefont{Wootters}(1998)}]{wootters98}
    \bibinfo{author}{\bibfnamefont{W.~K.} \bibnamefont{Wootters}},
  \bibinfo{journal}{Phys. Rev. Lett.} \textbf{\bibinfo{volume}{80}},
  \bibinfo{pages}{2245} (\bibinfo{year}{1998}).

\bibitem[{\citenamefont{Uhlmann}(2000)}]{uhlmann00}
    \bibinfo{author}{\bibfnamefont{A.}~\bibnamefont{Uhlmann}},
  \bibinfo{journal}{Phys. Rev. A} \textbf{\bibinfo{volume}{62}},
  \bibinfo{pages}{032307} (\bibinfo{year}{2000}).

\bibitem[{\citenamefont{Mintert et~al.}(2004)\citenamefont{Mintert, Ku\'s,
    and
  Buchleitner}}]{mkb04}
\bibinfo{author}{\bibfnamefont{F.}~\bibnamefont{Mintert}},
  \bibinfo{author}{\bibfnamefont{M.}~\bibnamefont{Ku\'s}}, \bibnamefont{and}
  \bibinfo{author}{\bibfnamefont{A.}~\bibnamefont{Buchleitner}},
  \bibinfo{journal}{Phys. Rev. Lett.} \textbf{\bibinfo{volume}{92}},
  \bibinfo{pages}{167902} (\bibinfo{year}{2004}).

\bibitem[{\citenamefont{Mintert
  et~al.}(2005{\natexlab{a}})\citenamefont{Mintert, Ku\'s, and
  Buchleitner}}]{mkb05}
\bibinfo{author}{\bibfnamefont{F.}~\bibnamefont{Mintert}},
  \bibinfo{author}{\bibfnamefont{M.}~\bibnamefont{Ku\'s}}, \bibnamefont{and}
  \bibinfo{author}{\bibfnamefont{A.}~\bibnamefont{Buchleitner}},
  \bibinfo{journal}{Phys. Rev. Lett.} \textbf{\bibinfo{volume}{95}},
  \bibinfo{pages}{260502} (\bibinfo{year}{2005}{\natexlab{a}}).

\bibitem[{\citenamefont{Mintert
  et~al.}(2005{\natexlab{b}})\citenamefont{Mintert, Carvalho, Ku\'s, and
  Buchleitner}}]{mckb05}
\bibinfo{author}{\bibfnamefont{F.}~\bibnamefont{Mintert}},
  \bibinfo{author}{\bibfnamefont{A.~R.~R.} \bibnamefont{Carvalho}},
  \bibinfo{author}{\bibfnamefont{M.}~\bibnamefont{Ku\'s}}, \bibnamefont{and}
  \bibinfo{author}{\bibfnamefont{A.}~\bibnamefont{Buchleitner}},
  \bibinfo{journal}{Phys. Rep.} \textbf{\bibinfo{volume}{415}},
  \bibinfo{pages}{207–} (\bibinfo{year}{2005}{\natexlab{b}}).

\bibitem[{\citenamefont{Mintert}(2007{\natexlab{a}})}]{mintert07}
    \bibinfo{author}{\bibfnamefont{F.}~\bibnamefont{Mintert}},
  \bibinfo{journal}{Phys. Rev. A} \textbf{\bibinfo{volume}{75}},
  \bibinfo{eid}{052302} (\bibinfo{year}{2007}{\natexlab{a}}).

\bibitem[{\citenamefont{Mintert and Buchleitner}(2007)}]{mintert07a}
    \bibinfo{author}{\bibfnamefont{F.}~\bibnamefont{Mintert}}
    \bibnamefont{and}
  \bibinfo{author}{\bibfnamefont{A.}~\bibnamefont{Buchleitner}},
  \bibinfo{journal}{Phys. Rev. Lett.} \textbf{\bibinfo{volume}{98}},
  \bibinfo{eid}{140505} (\bibinfo{year}{2007}).

\bibitem[{\citenamefont{Mintert}(2007{\natexlab{b}})}]{mintert07b}
    \bibinfo{author}{\bibfnamefont{F.}~\bibnamefont{Mintert}},
  \bibinfo{journal}{Appl. Phys. B} \textbf{\bibinfo{volume}{89}},
  \bibinfo{pages}{493 } (\bibinfo{year}{2007}{\natexlab{b}}).

\bibitem[{\citenamefont{Aolita and Mintert}(2006)}]{aolita06}
    \bibinfo{author}{\bibfnamefont{L.}~\bibnamefont{Aolita}}
    \bibnamefont{and}
  \bibinfo{author}{\bibfnamefont{F.}~\bibnamefont{Mintert}},
  \bibinfo{journal}{Phys. Rev. Lett.} \textbf{\bibinfo{volume}{97}},
  \bibinfo{eid}{050501} (\bibinfo{year}{2006}).

\bibitem[{\citenamefont{Aolita et~al.}(2008)\citenamefont{Aolita,
    Buchleitner,
  and Mintert}}]{aolita08}
\bibinfo{author}{\bibfnamefont{L.}~\bibnamefont{Aolita}},
  \bibinfo{author}{\bibfnamefont{A.}~\bibnamefont{Buchleitner}},
  \bibnamefont{and} \bibinfo{author}{\bibfnamefont{F.}~\bibnamefont{Mintert}},
  \bibinfo{journal}{Phys. Rev. A} \textbf{\bibinfo{volume}{78}},
  \bibinfo{eid}{022308} (pages~\bibinfo{numpages}{4}) (\bibinfo{year}{2008}).

\bibitem[{\citenamefont{Zhang et~al.}(2008)\citenamefont{Zhang, Gong, Zhang,
  and Guo}}]{zhang08a}
\bibinfo{author}{\bibfnamefont{C.-J.} \bibnamefont{Zhang}},
  \bibinfo{author}{\bibfnamefont{Y.-X.} \bibnamefont{Gong}},
  \bibinfo{author}{\bibfnamefont{Y.-S.} \bibnamefont{Zhang}}, \bibnamefont{and}
  \bibinfo{author}{\bibfnamefont{G.-C.} \bibnamefont{Guo}},
  \bibinfo{journal}{Phys. Rev. A} \textbf{\bibinfo{volume}{78}},
  \bibinfo{eid}{042308} (\bibinfo{year}{2008}).

\bibitem[{\citenamefont{Horodecki et~al.}(2009)\citenamefont{Horodecki,
  Horodecki, Horodecki, and Horodecki}}]{horodecki09}
\bibinfo{author}{\bibfnamefont{R.}~\bibnamefont{Horodecki}},
  \bibinfo{author}{\bibfnamefont{P.}~\bibnamefont{Horodecki}},
  \bibinfo{author}{\bibfnamefont{M.}~\bibnamefont{Horodecki}},
  \bibnamefont{and}
  \bibinfo{author}{\bibfnamefont{K.}~\bibnamefont{Horodecki}},
  \bibinfo{journal}{Rev. Mod. Phys.} \textbf{\bibinfo{volume}{81}},
  \bibinfo{eid}{865} (\bibinfo{year}{2009}).

\bibitem[{\citenamefont{Schliemann
  et~al.}(2001{\natexlab{a}})\citenamefont{Schliemann, Loss, and
  MacDonald}}]{schliemann01}
\bibinfo{author}{\bibfnamefont{J.}~\bibnamefont{Schliemann}},
  \bibinfo{author}{\bibfnamefont{D.}~\bibnamefont{Loss}}, \bibnamefont{and}
  \bibinfo{author}{\bibfnamefont{A.~H.} \bibnamefont{MacDonald}},
  \bibinfo{journal}{Phys. Rev. B} \textbf{\bibinfo{volume}{63}},
  \bibinfo{pages}{085311} (\bibinfo{year}{2001}{\natexlab{a}}).

\bibitem[{\citenamefont{Schliemann
  et~al.}(2001{\natexlab{b}})\citenamefont{Schliemann, Cirac, Ku\'s,
  Lewenstein, and Loss}}]{sckll01}
\bibinfo{author}{\bibfnamefont{J.}~\bibnamefont{Schliemann}},
  \bibinfo{author}{\bibfnamefont{J.~I.} \bibnamefont{Cirac}},
  \bibinfo{author}{\bibfnamefont{M.}~\bibnamefont{Ku\'s}},
  \bibinfo{author}{\bibfnamefont{M.}~\bibnamefont{Lewenstein}},
  \bibnamefont{and} \bibinfo{author}{\bibfnamefont{D.}~\bibnamefont{Loss}},
  \bibinfo{journal}{Phys. Rev. A} \textbf{\bibinfo{volume}{64}},
  \bibinfo{pages}{022303} (\bibinfo{year}{2001}{\natexlab{b}}).

\bibitem[{\citenamefont{Horn and Johnson}(1985)}]{horn85}
    \bibinfo{author}{\bibfnamefont{R.~A.} \bibnamefont{Horn}}
    \bibnamefont{and}
  \bibinfo{author}{\bibfnamefont{C.~R.} \bibnamefont{Johnson}},
  \emph{\bibinfo{title}{{M}atrix {A}nalysis}} (\bibinfo{publisher}{Cambridge
  University Press}, \bibinfo{address}{Cambridge}, \bibinfo{year}{1985}).

\bibitem[{\citenamefont{Herbut and Vujicic}(1987)}]{herbut87}
    \bibinfo{author}{\bibfnamefont{F.}~\bibnamefont{Herbut}}
    \bibnamefont{and}
  \bibinfo{author}{\bibfnamefont{M.}~\bibnamefont{Vujicic}},
  \bibinfo{journal}{J. Phys. A} \textbf{\bibinfo{volume}{20}},
  \bibinfo{pages}{5555} (\bibinfo{year}{1987}).

\bibitem[{\citenamefont{Grobe et~al.}(1994)\citenamefont{Grobe,
  Rz\c{a}\.zewski, and Eberly}}]{grobe94}
\bibinfo{author}{\bibfnamefont{R.}~\bibnamefont{Grobe}},
  \bibinfo{author}{\bibfnamefont{K.}~\bibnamefont{Rz\c{a}\.zewski}},
  \bibnamefont{and} \bibinfo{author}{\bibfnamefont{J.~H.}
  \bibnamefont{Eberly}}, \bibinfo{journal}{J. Phys. B}
  \textbf{\bibinfo{volume}{27}}, \bibinfo{pages}{L503} (\bibinfo{year}{1994}).

\bibitem[{\citenamefont{Eckert et~al.}(2002)\citenamefont{Eckert, Schliemann,
  Bru{\ss}, and Lewenstein}}]{eckert02}
\bibinfo{author}{\bibfnamefont{K.}~\bibnamefont{Eckert}},
  \bibinfo{author}{\bibfnamefont{J.}~\bibnamefont{Schliemann}},
  \bibinfo{author}{\bibfnamefont{D.}~\bibnamefont{Bru{\ss}}}, \bibnamefont{and}
  \bibinfo{author}{\bibfnamefont{M.}~\bibnamefont{Lewenstein}},
  \bibinfo{journal}{Ann. Phys.} \textbf{\bibinfo{volume}{299}},
  \bibinfo{pages}{88} (\bibinfo{year}{2002}).

\bibitem[{\citenamefont{Li et~al.}(2001)\citenamefont{Li, Zeng, Liu, and
  Long}}]{li01}
\bibinfo{author}{\bibfnamefont{Y.~S.} \bibnamefont{Li}},
  \bibinfo{author}{\bibfnamefont{B.}~\bibnamefont{Zeng}},
  \bibinfo{author}{\bibfnamefont{X.~S.} \bibnamefont{Liu}}, \bibnamefont{and}
  \bibinfo{author}{\bibfnamefont{G.~L.} \bibnamefont{Long}},
  \bibinfo{journal}{Phys. Rev. A} \textbf{\bibinfo{volume}{64}},
  \bibinfo{pages}{054302} (\bibinfo{year}{2001}).

\bibitem[{\citenamefont{Ghirardi et~al.}(2002)\citenamefont{Ghirardi,
  Marinatto, and Weber}}]{ghirardi02}
\bibinfo{author}{\bibfnamefont{G.}~\bibnamefont{Ghirardi}},
  \bibinfo{author}{\bibfnamefont{L.}~\bibnamefont{Marinatto}},
  \bibnamefont{and} \bibinfo{author}{\bibfnamefont{T.}~\bibnamefont{Weber}},
  \bibinfo{journal}{J. Stat. Phys.} \textbf{\bibinfo{volume}{108}},
  \bibinfo{pages}{49} (\bibinfo{year}{2002}).

\bibitem[{\citenamefont{Pa\v{s}kauskas and You}(2001)}]{paskauskas01}
    \bibinfo{author}{\bibfnamefont{R.}~\bibnamefont{Pa\v{s}kauskas}}
  \bibnamefont{and} \bibinfo{author}{\bibfnamefont{L.}~\bibnamefont{You}},
  \bibinfo{journal}{Phys. Rev. A} \textbf{\bibinfo{volume}{64}},
  \bibinfo{pages}{042310} (\bibinfo{year}{2001}).

\bibitem[{\citenamefont{Ghirardi and Marinatto}(2004)}]{ghirardi04}
    \bibinfo{author}{\bibfnamefont{G.}~\bibnamefont{Ghirardi}}
    \bibnamefont{and}
  \bibinfo{author}{\bibfnamefont{L.}~\bibnamefont{Marinatto}},
  \bibinfo{journal}{Phys. Rev. A} \textbf{\bibinfo{volume}{70}},
  \bibinfo{pages}{012109} (\bibinfo{year}{2004}).

\bibitem[{\citenamefont{Herbut}(2001)}]{herbut01}
    \bibinfo{author}{\bibfnamefont{F.}~\bibnamefont{Herbut}},
    \bibinfo{journal}{Am.
  J. Phys.} \textbf{\bibinfo{volume}{69}}, \bibinfo{pages}{207}
  (\bibinfo{year}{2001}).

\bibitem[{\citenamefont{Tichy et~al.}()\citenamefont{Tichy, de~Melo, Ku\'s,
  Mintert, and Buchleitner}}]{tmkb-preprint}
\bibinfo{author}{\bibfnamefont{M.~C.} \bibnamefont{Tichy}},
  \bibinfo{author}{\bibfnamefont{F.}~\bibnamefont{de~Melo}},
  \bibinfo{author}{\bibfnamefont{M.}~\bibnamefont{Ku\'s}},
  \bibinfo{author}{\bibfnamefont{F.}~\bibnamefont{Mintert}}, \bibnamefont{and}
  \bibinfo{author}{\bibfnamefont{A.}~\bibnamefont{Buchleitner}},
  \bibinfo{note}{arxiv-0902.1684}.

\bibitem[{\citenamefont{Hall}(2003)}]{hall03}
    \bibinfo{author}{\bibfnamefont{B.}~\bibnamefont{Hall}},
  \emph{\bibinfo{title}{{L}ie groups, {L}ie algebras, and representations: an
  elementary introduction}} (\bibinfo{publisher}{Springer},
  \bibinfo{address}{New York}, \bibinfo{year}{2003}).

\bibitem[{\citenamefont{Lichtenstein}(1982)}]{lichtenstein82}
    \bibinfo{author}{\bibfnamefont{W.}~\bibnamefont{Lichtenstein}},
  \bibinfo{journal}{Proc. Am. Math. Soc.} \textbf{\bibinfo{volume}{84}},
  \bibinfo{pages}{605} (\bibinfo{year}{1982}).

\bibitem[{\citenamefont{Barut and R\c{a}czka}(1980)}]{barut80}
    \bibinfo{author}{\bibfnamefont{A.~O.} \bibnamefont{Barut}}
    \bibnamefont{and}
  \bibinfo{author}{\bibfnamefont{R.}~\bibnamefont{R\c{a}czka}},
  \emph{\bibinfo{title}{{T}heory of group representations and applications}}
  (\bibinfo{publisher}{PWN}, \bibinfo{address}{Warszawa},
  \bibinfo{year}{1980}).

\bibitem[{\citenamefont{Jamio{\l}kowski}(1972)}]{jamiolkowski72}
    \bibinfo{author}{\bibfnamefont{A.}~\bibnamefont{Jamio{\l}kowski}},
  \bibinfo{journal}{Rep. Math. Phys.} \textbf{\bibinfo{volume}{3}},
  \bibinfo{pages}{275} (\bibinfo{year}{1972}).

\bibitem[{\citenamefont{Ku\'s and Bengtsson}(2009)}]{kb09}
    \bibinfo{author}{\bibfnamefont{M.}~\bibnamefont{Ku\'s}} \bibnamefont{and}
  \bibinfo{author}{\bibfnamefont{I.}~\bibnamefont{Bengtsson}},
  \bibinfo{journal}{Phys. Rev. A} \textbf{\bibinfo{volume}{80}},
  \bibinfo{eid}{022319} (\bibinfo{year}{2009}).

\bibitem[{\citenamefont{Hughston et~al.}(1993)\citenamefont{Hughston, Jozsa,
  and Wootters}}]{hughston93}
\bibinfo{author}{\bibfnamefont{L.~P.} \bibnamefont{Hughston}},
  \bibinfo{author}{\bibfnamefont{R.}~\bibnamefont{Jozsa}}, \bibnamefont{and}
  \bibinfo{author}{\bibfnamefont{W.~K.} \bibnamefont{Wootters}},
  \bibinfo{journal}{Phys. Lett. A} \textbf{\bibinfo{volume}{183}},
  \bibinfo{pages}{14} (\bibinfo{year}{1993}).

\end{thebibliography}

\end{document}